\documentclass[letterpaper,12pt]{article}
\usepackage{amsmath}
\usepackage{amsfonts}
\usepackage{amssymb}
\usepackage{graphicx}
\usepackage{physics}
\usepackage{latexsym}
\usepackage{epsfig}
\usepackage{pstricks}
\usepackage{stmaryrd}
\usepackage{rotating}
\usepackage[english]{babel}
\usepackage{setspace}
\usepackage[utf8]{inputenc}
\usepackage{natbib}
\usepackage[T1]{fontenc}
\usepackage{lmodern}
\usepackage{enumitem}
\usepackage{csquotes}
\usepackage{amsthm}
\usepackage{xcolor}
\usepackage[margin=1in]{geometry}
\begin{document}
\begin{center}	
\begin{Large}
\textbf{Reassessing the strength of a class of Wigner's friend no-go theorems}\\
\end{Large}
\end{center}

\begin{center}
\begin{large}
Elias Okon\\
\end{large}
\textit{Universidad Nacional Aut\'onoma de M\'exico, Mexico City, Mexico.}\\[1cm]
\end{center}

Two recent, prominent theorems---the ``no-go theorem for observer-independent facts'' and the ``Local Friendliness no-go theorem''---employ so-called extended Wigner's friend scenarios to try to impose novel, non-trivial constraints on the possible nature of physical reality. While the former is argued to entail that there can be no theory in which the results of Wigner and his friend can both be considered objective, the latter is said to place on reality stronger constraints than the Bell and Kochen-Specker theorems. Here, I conduct a thorough analysis of these theorems and show that they suffer from a list of shortcomings that question their validity and limit their strength. I conclude that the ``no-go theorem for observer-independent facts'' and the ``Local Friendliness no-go theorem'' fail to impose significant constraints on the nature of physical reality.

\onehalfspacing

\section{Introduction}

Recent years have seen a renewed interest in the Wigner's friend paradox. In particular, so-called extended Wigner's friend scenarios (EWFS) have been in vogue as a means to display limits regarding the level of \emph{objectivity} one can make use of while interpreting quantum theory. These EWFS consist of settings in which entangled systems are shared by sets of Wigner’s friend arrangements; that is, they are made up of Bell-type experimental settings, in which the components of entangled systems are sent to different labs, but with such labs enhanced with Wigner’s friend arrangements, where the friend first measures the received component and then a `superobserver' measures the whole lab.

The current surge in interest in EWFS has been somehow unsystematic, making it difficult to assess the real value of the different efforts and the relations among them. One useful way to classify them is as follows. On the one hand, there are those proposals which leave fixed the measurement settings of all observers involved, both friends and superobservers. The idea is to design the experiments in such a way that one employs, on a single run, measurement settings that, in a standard Bell-type scenario, would correspond to different possibilities. One then tries to use the correlations among the results of all observers to set limits on possible interpretations of quantum theory. Efforts in this direction include the Pusey-Masanes \citep{Pusey} and Żukowski-Markiewicz \citep{Zuk} arguments, which try to challenge the objectivity of quantum measurement outcomes, and the Frauchiger-Renner theorem \citep{FR}, which allegedly shows that quantum theory does not apply at all scales. The problem with these arguments, though, is that they rely on a mistaken assumption regarding the correlations between the results of the friends and those of superobservers, rendering them invalid (see \cite{Obj,lazarovici} for details).

On the other hand, there are proposals which, as in the standard Bell scenario, do allow for the observers to choose what measurement to perform. In particular, in these settings, on each run of the experiment, each superobserver only performs one of a list of possible measurements. Moreover, these proposals build their results only employing correlations among the results of the superobservers. This group includes the ``no-go theorem for observer-independent facts'' in \cite{Bru1}, as well as the ``Local Friendliness no-go theorem'' in \cite{LF} (see also \cite{LFp}). These theorems are the main subject of this assessment.

Building on \cite{Bru2}, the first (to my knowledge) to employ an EWFS, \cite{Bru1} presents a ``no-go theorem for observer-independent facts''. To do so, he considers two Wigner’s friend arrangements, sharing an entangled pair, and a particular sequence of measurements performed by the friends and the superobservers. Then, by positing four assumptions---`Universal validity of quantum theory', `Locality', `Freedom of choice' and `Observer-independent facts'---he derives a theorem interpreted as showing that there can be no theory in which the results of both the superobservers and the friends can jointly be considered as objective properties of the world.

\cite{LF}, however, argues that Brukner's `Observer-independent facts' assumption is equivalent to Kochen-Specker non-contextuality (KSNC) and that Brukner's result can be obtained from `Freedom of choice' and KSNC alone (no `Locality' required). Moreover, by arguing that the Kochen-Specker theorem already establishes that `Freedom of choice' and KSNC lead to a contradiction with quantum mechanics, it is concluded that Brukner's theorem fails to set constraints on the objectivity of observations. Nevertheless, \cite{LF} maintains that there is something of value in Brukner's arrangement and that, in fact, it can be used to derive a theorem by retaining his `Freedom of choice' and `Locality' assumptions, but replacing his `Observer-independent facts' by what they call the `Absoluteness of Observed Events' assumption. Since it is argued that the new set of assumptions is strictly weaker than those of Bell and Kochen-Specker theorems, the new result is advertised as placing strictly stronger constraints on physical reality than these old theorems.

Here, I perform a thorough analysis of the ``no-go theorem for observer-independent facts'' of \cite{Bru1} and the ``Local Friendliness no-go theorem'' of \cite{LF}. Regarding the former, I show that it depends on a non-trivial hidden assumption and that it relies on a mistaken application of standard quantum mechanics to the EWFS considered. As for the latter, I argue that it is not true that its assumptions are strictly weaker than those of Bell and that it relies on the same mistaken application of standard quantum mechanics. I conclude that the theorems fail to set interesting constraints on the possible nature of physical reality.

My manuscript is organized as follows. In section \ref{TT}, I present an overview of the ``no-go theorem for observer-independent facts'' of \cite{Bru1} and the ``Local Friendliness no-go theorem'' of \cite{LF}. Then, in section \ref{MA}, I offer my assessment of such theorems. Finally, in section \ref{Co}, I state my conclusions.

\section{The theorems}
\label{TT}

We start with an overview of the ``no-go theorem for observer-independent facts'' in \cite{Bru1} and the ``Local Friendliness no-go theorem'' in \cite{LF}.

\subsection{The no-go theorem for observer-independent facts}

Elaborating on \cite{Bru2}---the first to consider a Bell-type scenario supplemented with Wigner’s friend arrangements---\cite{Bru1} asks whether there exists a theory, possibly different from quantum theory, in which the results of both Wigner and his friend can jointly be considered observer-independent, objective facts of the world. The question is given a negative answer with the introduction of a ``no-go theorem for observer-independent facts''.

To construct the theorem, Brukner considers an EWFS with two Wigner’s friend \linebreak arrangements---lab 1 with Charlie inside and Alice outside and lab 2 with Debbie inside and Bob outside---each receiving a particle of an entangled pair. Inside their labs, Charlie and Debbie perform spin measurements along direction $z$. Let's denote by $\ket{Z_\pm}_i $ the state of lab $i$ after the friend measures the spin along $z$ and finds the result $\pm$. Then, Alice and Bob perform, on their respective labs, one of two measurements: a ``Z measurement'' in basis $\{ \ket{Z_+}_i,\ket{Z_-}_i\}$ or an ``X measurement'' in basis $\{ \ket{X_+}_i = \frac{1}{\sqrt{2}}\left( \ket{Z_+}_i+\ket{Z_-}_i \right), \ket{X_-}_i = \frac{1}{\sqrt{2}}\left( \ket{Z_+}_i - \ket{Z_-}_i \right) \}$. That is, on each run of the experiment, each superobserver only performs one of these two possible measurements.

The setting described above is used by Brukner to argue for the mutual incompatibility of four assumptions: (1) `Universal validity of quantum theory', establishing that quantum predictions hold at any scale; (2) `Locality', requiring that the choice of measurement on one side has no influence on the outcomes of the other; (3) `Freedom of choice', stipulating statistical independence between measurement settings and the rest of the experiment and (4) `Observer-independent facts', demanding for there to exist a joint assignment of truth values to propositions about the outcomes of all measurements. In particular, according to Brukner, this last assumption implies that the statements he calls $A_1$: ``the pointer of Charlie's apparatus points to result $+$'' and $A_2$: ``after performing an X measurement, the pointer of Alice's apparatus points to result $X_-$'', have well-defined truth values. That is, all such statements are assumed to have truth values, regardless of what measurements the superobservers decide to perform.

The incompatibility among Brukner's assumptions is shown as follows. He denotes by $A_1$ and $A_2$ the two measurement settings of Alice, which correspond to the observational statements Charlie and Alice can make about their respective outcomes (analogously with $B_1$ and $B_2$ for Bob and Debbie). Then, it is noted that assumptions (2), (3) and (4) imply the existence of local hidden variables, which determine $A_1$, $A_2$, $B_1$ and $B_2$ (with values $\pm1$). Moreover, it is noted that these assumptions imply the existence of a joint distribution, $p(A_1, A_2, B_1, B_2)$, whose marginals (e.g., $p(A_1,B_2) = \sum_{A_2,B_1} p(A_1, A_2, B_1, B_2))$ satisfy the CHSH inequality: $E(A_1,B_1)+E(A_1,B_2)+E(A_2,B_1)-E(A_2,B_2) \le 2$ (with, e.g., $E(A_1,B_2) = \sum_{A_1,B_2} A_1 B_2 p(A_1,B_2))$. The problem is that there exist entangled states to be distributed among the two labs, such as
\begin{equation}
\label{psi}
 \frac{1}{2} \left[ \sin(\pi/8) \ket{+_z,+_z} + \cos(\pi/8) \ket{+_z, -_z} - \cos(\pi/8) \ket{-_z, +_z} + \sin(\pi/8) \ket{-_z, -_z} \right] ,
\end{equation}
which lead to quantum predictions that violate the CHSH inequality. Since the assumptions are shown to be inconsistent, the conclusion reached by Brukner is that there can be no theoretical framework where the results of different observers can jointly be considered objective facts of the world.

\subsection{The Local Friendliness no-go theorem}
\label{LF}

The starting point of the ``Local Friendliness no-go theorem'' presented in \cite{LF} (see also \citet[section 4]{LFp}), is an analysis of Brukner's ``no-go theorem for observer-independent facts''. According to \cite{LF}, Brukner's `Observer-independent facts' assumption demands propositions about all observables \emph{that might be measured}, by an observer or a superobserver, to be assigned a truth value, independently of which measurements the superobservers perform. That is, they read Brukner's assumption as implying, for instance, that even when Alice performs a Z measurement, statements regarding the result of Alice's \emph{unperformed} X measurement have well-defined truth values.

Given this reading, they argue that Brukner's `Observer-independent facts' assumption is, in fact, equivalent to Kochen-Specker non-contextuality (KSNC). Moreover, they argue that such an assumption, together with `Freedom of choice', are sufficient to derive Brukner's result. Since the Kochen-Specker theorem already establishes that `Freedom of choice' and KSNC are incompatible with quantum mechanics, they conclude that the `Observer-independent facts' assumption plays no role in the theorem so, in the end, it fails in its mission to place limits on the objectivity of facts.\footnote{\cite{Healey} also interprets Brukner's `Observer-independent facts' assumption in terms of counterfactual measurements and arrives at a similar conclusion as \cite{LF} regarding the import of Brukner's theorem.} 

Still, according to \cite{LF}, Brukner's arrangement is valuable and, in fact, can be used to derive a strong result. To do so,  echoing Brukner, they consider two perfectly isolated vaults, one containing Charlie and the other Debbie, each receiving a particle of an entangled pair. Charlie and Debbie then perform spin measurements on a fixed basis, obtaining outcomes $C$ and $D$. Then, Alice and Bob perform measurements $X$ and $Y$, selected out of a list of possible measurements, on Charlie's and Debbie's vaults, respectively. Their outcomes are denoted $A$ and $B$. The possible measurements by Alice are described as follows. If $X=1$, Alice opens Charlie's vault, asks him what he observed and sets her result equal to that of Charlie. If $X \neq 1$, Alice performs a measurement on the contents of the whole vault, including Charlie, in a basis incompatible with the one used by Charlie. This can be done, it is argued, by reversing the unitary evolution that entangled Charlie and his apparatus with his particle, and performing a measurement on the particle alone on a different basis (the possible measurements by Bob are obtained by substituting $X$ for $Y$ and Charlie for Debbie).

To analyze this setup, \cite{LF} introduces three assumptions: `Absoluteness of Observed Events' (AOE), `No-Superdeterminism' (NSD) and `Locality' (L). The conjunction of these three assumptions is then called `Local Friendliness' (LF). These assumptions are described as follows.

The idea behind AOE is that every observed event exists absolutely, not relatively to an observer, system or context. That is, that performed experiments have observer-independent, absolute results. For the experiment under consideration, this implies that, in each run, there exists a well-defined value for the result observed by each observer; i.e., in each run, there is a well-defined value for $A$, $B$, $C$ and $D$. This, in turn, implies the existence of a joint distribution, $P(ABCD|XY)$, from which the joint probability for the results of Alice and Bob, $P(AB|XY)$, can be obtained. Moreover, the joint distribution must ensure consistency between the outcomes of friends and superobservers when $X,Y =1$. Given all this, for the experiment considered, AOE is formalized as follows:
\begin{itemize}
\item $\exists $ P(ABCD|XY) such that
\begin{description}
\item[i.] $P(AB|XY) = \sum_{C,D} P(ABCD|XY) \quad \forall \quad A, B, X, Y$
\item[ii.] $P(A|CD,X=1,Y)= \delta_{AC} \quad \forall \quad A, C, D, Y$
\item[iii.] $P(B|CD,X,Y=1)= \delta_{BD} \quad \forall \quad B, C, D, X$ .
\end{description}
\end{itemize}

The motivation for NSD is the idea that the experimental settings can be chosen freely. It is argued to be a formalization of the `Freedom of choice' assumption used to derive Bell's theorem, and is stated as follows: ``any set of events on a space-like hypersurface is uncorrelated with any set of freely chosen actions subsequent to that space-like hypersurface''. In the experiment under consideration, and assuming AOE, NSD is taken to imply that $C$ and $D$ are independent of the choices $X$ and $Y$; that is:
\begin{itemize}
\item $P(CD|XY) = P(CD) \quad \forall \quad C, D, X, Y$ .
\end{itemize}

Finally, L prohibits the influence of local settings on distant outcomes. That is, it is the assumption usually called `Parameter independence'. In \cite{LF}, it is enunciated as follows: ``the probability of an observable event $e$ is unchanged by conditioning on a space-like-separated free choice $z$, even if it is already conditioned on other events not in the future light-cone of $z$''. For the experiment in question, and assuming AOE, L is said to imply:
\begin{itemize}
\item $P(A|CDXY) = P(A|CDX) \quad \forall \quad A, C, D, X, Y$
\item $P(B|CDXY) = P(B|CDY) \quad \forall \quad B, C, D, X, Y$ .
\end{itemize}

The construction of the theorem then proceeds in two steps. On the one hand, it is shown that the LF assumptions imply a set of constraints on the joint distribution for the results of Alice and Bob, $P(AB|XY)$. Such constraints are expressed in the form of the ``LF inequalities''. On the other hand, it is argued that if a superobserver can perform arbitrary quantum operations on an observer and its environment---that is, if quantum evolution (including quantum measurement) is controllable on such a scale---then quantum mechanics predicts the violation of the LF inequalities. In other words, if the quantum operations required by the experiment can, in principle, be performed, then quantum mechanics predicts the violation of the LF inequalities.

With all this in mind, by defining a `physical theory' as ``any theory that correctly predicts the correlations between the outcomes observed by the superobservers Alice and Bob'', the LF theorem is stated in \cite{LF} as follows:
\begin{quotation}
\noindent \textbf{LF Theorem}: If a superobserver can perform arbitrary quantum operations on an observer and its environment, then no physical theory can satisfy the LF assumptions.
\end{quotation}

Regarding the LF inequalities, it is noted that, for the specific experiment considered by Brukner---in which the superobservers only have two binary-outcome measurement options---the set of LF correlations is identical to those allowed by Bell. This, however, is not the case in general. In more complicated scenarios, in which, for instance, the superobservers have more than two measurement choices or measurements with more than two possible outcomes, the set of LF correlations can be larger than the set of Bell correlations. In fact, it is argued that, for a given scenario, the set of LF correlations strictly contains the set of Bell correlations. This implies that, for a given scenario, it is possible for quantum correlations to violate a Bell inequality, while satisfying all of the LF inequalities (but not the other way around).

An important claim in \cite{LF} is that their theorem ``places strictly stronger constraints on physical reality than Bell's theorem''. To reach such a conclusion, they note that the derivation of the Bell inequalities requires NSD and `Factorizability'. The latter, in turn, follows from the conjunction of `Parameter independence' (here called L) and `Outcome independence'. Moreover, they argue that the AOE assumption is necessary, although often left implicit, for the derivation of Bell's theorem. From this, they conclude that the LF assumptions are strictly weaker than the assumptions needed to derive Bell's theorem, which means that a violation of the LF inequalities would have stronger implications than violations of the Bell inequalities. Finally, it is pointed out that, as Bell's, the LF theorem is theory-independent, in the sense that its conclusions hold for any theory, as long as the quantum predictions are realized in the laboratory. 

Regarding the realization of such a test, \cite{LF} reports the violation of the LF inequalities in an experiment in which the paths of photons play the role of observers. It is readily acknowledged that such an experiment is simply a proof of concept, and that a true test of the inequalities, with observers of adequate complexity, is well beyond present technology. However, it is argued that, if the friend were an artificial intelligence algorithm, simulated in a quantum computer, then there would be good reason to think that quantum mechanics would allow control of the type required.

\section{Assessing the theorems}
\label{MA}

I am finally in position to explore these theorems in detail. I start with an evaluation of Brukner's work.

\subsection{Evaluation of the no-go theorem for observer-independent facts}

The initial observation I make is that the discussion in \cite{Bru1} of the key `Observer-independent facts' assumption is somehow ambiguous and allows for two different interpretations. The issue is that it is not clear whether what Brukner calls the observational statement $A_1$ (``The pointer of Wigner's friend's apparatus points to result $z+$''), refers to an observational statement uttered by Charlie, or by Alice when she performs a Z measurement. This is of relevance because these different possible readings lead to different contents for the `Observer-independent facts' assumption. 

One option is to take $A_1$ as a statement uttered by Alice. In that case, since the `Observer-independent facts' assumption requires ``an assignment of truth values to statements $A_1$ and $A_2$ independently of which measurement Wigner performs'', then the assumption would demand, not only for all performed experiments to have well-defined results, but also for (at least some) unperformed experiments to do so. This seems to be the interpretation of the `Observer-independent facts' assumption offered in \cite{LF} where, as we saw, it is taken as demanding propositions about all observables \emph{that might be measured} to be assigned a truth value, independently of whether they were actually performed or not (\cite{Healey} contains the same interpretation). That is, they read Brukner's assumption as implying that, even when Alice performs a Z measurement, a statement regarding the result of her \emph{unperformed} X measurement has to have a well-defined value. This is why, in \cite{LF}, it is concluded that the `Observer-independent facts' assumption is equivalent to KSNC. 

However, even with this reading of Brukner's assumption, it seems to me that this identification is illegitimate. KSNC is an assumption demanding for the results of all experiments to be independent of the \emph{context} of the measurement and, in particular, independent of which other measurements were performed simultaneously. I fail to see how, by reading $A_1$ as a statement by Alice, the `Observer-independent facts' is equivalent to demanding non-contextuality.\footnote{That is, KSNC would demand, for instance, for the result of Alice's experiment to be independent of what Bob decides to measure but, by reading $A_1$ as a statement by Alice, the `Observer-independent facts' demands, say, for the result of Alice's $A_1$ measurement to be well-defined, even if such a measurement is not performed and Alice measures $A_2$. Moreover, the KSNC is a demand for all results, of all possible experiments. It is not clear to me that the particular demand by Brukner, made for a specific experiment, could straightforwardly be read as setting such a strong demand for all possible observables, in all possible experiments. In any case, it is important to clearly distinguish between a demand for experiments to yield observer-independent, absolute results---the motivation behind `Observer-independent facts'---and a demand for all measurements to passively, context-independently, reveal previously possessed well-defined values---the motivation behind KSNC. Of course, the latter is a much stronger requirement, already ruled out by the Kochen-Specker theorem.} At any rate, I agree with \cite{LF} that, by reading the `Observer-independent facts' assumption this way, Brukner's no-go theorem fails to set relevant limits on the objectivity of facts. The point is that, by interpreting $A_1$ as a statement by Alice (and $B_1$ as a statement by Bob), Brukner, at best, sets constraints on frameworks demanding objectivity, not only of performed experiments, but also of unperformed ones. It seems clear to me that such a demand goes well beyond what is usually required of a framework which seeks to maintain objectivity, so letting go of such a strong demand does not seem problematic at all.

What about opting for the other interpretation of $A_1$, namely, taking $A_1$ as a statement uttered by Charlie? This is by far the most natural reading of Brukner's `Observer-independent facts' assumption and the one in line with Brukner's own conclusions. As we saw, he takes the theorem to show that there can be no  framework where the results of Wigner and his friend can jointly be considered objective facts of the world. Therefore, it seems clear that what he has in mind is that the observational statement made by Charlie has a well-defined truth value, even if Alice decides to perform a Z measurement. Let's explore the import of Brukner's theorem, given this interpretation of $A_1$.

The starting point in the construction of the theorem is the observation that the `Observer-independent facts' assumption implies the existence of the joint distribution $p(A_1, A_2, B_1, B_2)$. It is important to note, though, that, given the interpretation of $A_1$ (and $B_1$) under consideration, this would mean the existence of a  joint distribution for the results of all four observers, Charlie, Alice, Debbie and Bob, in a run in which Alice and Bob decide to perform an X measurement---as opposed to a distribution for two possible results of Alice and two of Bob. That is, the `Observer-independent facts' assumption implies the reality of those particular four results of all four observers, so there must be a joint distribution for them. Since this is so, the marginals, expectation values and the CHSH inequality constructed out of $p(A_1, A_2, B_1, B_2)$ must also refer to these four particular measurements. 
The problem, though, is that Brukner proceeds to compare such results with the quantum predictions, not for those four measurements, but for the two possible results of Alice and the two possible results of Bob. That is, even though, at the beginning, he interprets $A_1$ and $B_1$ as observational statements made by Charlie and Debbie, in the end, he takes $A_1$ and $B_1$ to be the results of Alice and Bob, when they chose to perform a Z measurement.

There is, then, a \emph{hidden assumption} behind the derivation of the theorem in \cite{Bru1}, namely, that the results of Alice and Bob, when they perform a Z measurement, necessarily coincide with the results previously obtained by Charlie and Debbie, respectively.
In fact, for the construction of the theorem in \cite{LF}, such an assumption is made explicit (see items ii and iii of the AOE assumption). Therefore, I postpone a detailed discussion of the issue until I explore the LF theorem. For now, I just note the presence of this implicit, independent assumption in \cite{Bru1}, without which the theorem cannot be derived.

The next observation I make is related to the last step in the derivation of the theorem, namely, the comparison with quantum predictions. According to \cite{Bru1}, there are entangled states to be distributed among the two labs, such as that in Eq. (\ref{psi}), which lead to quantum predictions that violate the CHSH inequality. While this, of course, is the case for a standard Bell-type experiment, this does not automatically mean that this is also the case for the novel EWFS considered by Brukner, containing four agents and, in particular, two superobservers performing measurements over whole labs. One must be careful, then, to have in mind what is the correct \emph{physical interpretation} of the expectation values in a given CHSH inequality---and not because the quantum predictions violate such an inequality for a given experimental scenario, it means that they do so for others.

So what about the quantum predictions for Brukner's scenario? The problem is that, since it involves intermediate measurements by Charlie and Debbie, such predictions crucially depend on what exactly happens during measurements. The issue, of course, is that the standard (textbook) interpretation of quantum mechanics is hopelessly vague in that respect \citep{Bel:90}. The upshot is that \emph{standard quantum mechanics is simply unable to make concrete predictions for the scenario considered}---the measurement problem gets in the way. That is, the Wigner's friend scenarios under consideration are precisely the type of settings for which one cannot get away with pressing on, adopting an operational stance and ignoring the conceptual limitations of the standard framework.

It is often (implicitly or explicitly) assumed that a correct application of standard quantum mechanics to a Wigner's friend scenario means that Wigner must describe the laboratory, and all of its contents, as evolving unitarily---allegedly in accordance with the quantum rule for the evolution of isolated systems. This, presumably, is what Brukner has in mind with his `Universal validity of quantum theory' assumption. It is not clear to me, though, that standard quantum mechanics contains such a rule and, if so, that this would be a correct application of it. The issue is that standard quantum mechanics also contains the rule that systems collapse upon measurements by observers so, in a Wigner’s friend scenario, the unitary evolution of isolated systems and the collapse postulate collide. 

In any case, one could  simply \emph{stipulate} that the laboratory and all of its contents evolve unitarily, even during the measurements of the friends. The issue is that \cite{Bru1} also wants to assume that measurements yield objective results, but we know from \cite{Tim} that those two assumptions, unitarity and objective outcomes, are incompatible with another assumption inherent to the standard framework: that the physical description given by the quantum state is complete. Therefore, on pain of inconsistency, it is simply impossible to assume unitarity and objective outcomes, and to employ the standard framework to make predictions. One can, of course, employ a framework which solves the measurement problem to make predictions. The issue is that different frameworks---such as objective collapse models, pilot-wave theories or Everettian scenarios---lead to different predictions for the scenario in question. For instance, while pilot-wave does predict a violation of the CHSH inequality in Brukner's setting, objective collapse models do not (see \cite{Obj} for details).

Taking stock, the ``no-go theorem for observer-independent facts'' in \cite{Bru1} depends on an independent, hidden assumption demanding for the Z measurement results of Alice and Bob to coincide with those of Charlie and Debbie. Moreover, there is no such thing as the correct quantum predictions for the proposed experiment, so there is no quantum benchmark with which to compare the predictions of models satisfying the imposed constrains---namely, the four assumptions in \cite{Bru1} plus the hidden one mentioned above. That is, the constraints imposed on frameworks satisfying these assumptions are not really in conflict with quantum predictions.\footnote{A final comment regarding the conclusions in \cite{Bru1}. We saw that the theorem is interpreted as showing that there can be no theoretical framework where the results of different observers can jointly be considered objective facts of the world. However, even ignoring all the problems described above, and taking the result as valid, it is not clear why one would be forced to discard the `Observer-independent facts' assumption, as Brukner suggests. At best, the theorem would demand for (at least) one of its assumptions to be abandoned, without pointing to any one of them in particular.}  Below, I will show that the theorem in \cite{LF} suffers from quite similar issues. I delay a full assessment of the impact of these limitations on the validity, significance and strength of these theorems until after the detailed evaluation of the ``Local Friendliness no-go theorem'', to which we turn next.

\subsection{Evaluation of the Local Friendliness no-go theorem}

I start with a simple, intuitive proof of the LF inequality, for the particular case when Alice and Bob only have two binary-outcome measurement options---in which case the LF inequality coincides with the CHSH inequality. Consider the LF setting and an ensemble of runs for which Alice and Bob choose $X,Y=2$. By AOE, there is a joint distribution for the results of all observers and, by Fine's theorem \citep{Fine}, the expectation values of products of results, calculated with the marginals of such a joint distribution, satisfy the CHSH inequality
\begin{equation}
 \left\langle C_2 D_2 \right\rangle + \left\langle C_2 B_2 \right\rangle + \left\langle A_2 D_2 \right\rangle - \left\langle A_2 B_2 \right\rangle  \le 2
\end{equation}
with, e.g., $\left\langle C_2 B_2 \right\rangle$, the expectation value of the product of the results of Charlie and Bob, when $X,Y=2$.

Next, we employ the LF assumptions to transform this inequality, involving results of all four observers, into a CHSH inequality for the results of Alice and Bob. In particular, we note that 
\begin{equation}
\left\langle C_2 D_2 \right\rangle \overset{\text{NSD}}{=} \left\langle C_1 D_1 \right\rangle \overset{\text{AOE}}{=} \left\langle A_1 B_1 \right\rangle  
\end{equation}
and that
\begin{equation}
\left\langle C_2 B_2 \right\rangle \overset{\text{L,NSD}}{=}  \left\langle C_1 B_2 \right\rangle  \overset{\text{AOE}}{=}  \left\langle A_1 B_2 \right\rangle
\end{equation}
\begin{equation}
 \left\langle A_2 D_2 \right\rangle   \overset{\text{L,NSD}}{=}  \left\langle A_2 D_1 \right\rangle  \overset{\text{AOE}}{=}  \left\langle A_2 B_1 \right\rangle.
\end{equation}
That is, we see that NSD and L allow the identification of certain expectation values with the same observers, but different settings, and that AOE and, in particular, conditions ii and iii, allow us to substitute, under the right circumstances, Alice for Charlie and Bob for Debbie. Putting everything together,
\begin{equation}
 \left\langle A_1 B_1 \right\rangle + \left\langle A_1 B_2 \right\rangle + \left\langle A_2 B_1 \right\rangle - \left\langle A_2 B_2 \right\rangle  \le 2 ,
\end{equation}
that is, the CHSH inequality.\footnote{While standard proofs of Bell's theorem explicitly allow for the probabilities to change from run to run (through their dependence on what is usually denoted by $\lambda$), \cite{LF} does not allow for such a change---they explicitly take the results of the friends to play the role of hidden variables, but do not allow for additional variables that could alter the joint probability distribution from run to run. One can fix the problem by adding a dependency of the joint distributions on $\lambda$ and averaging over it. For this to work, one must assume, on top of the LF assumptions, the so-called `Settings Independence' assumption: $\rho(\lambda|X,Y) = \rho(\lambda)$. The upshot is that, given settings independence, it is enough for the expectation values calculated with the joint distribution of each possible value of $\lambda$ to satisfy the inequality, for the average of them, over $\lambda$, to do so.}


Above we saw that the starting point of the LF theorem is a critique of Brukner's `Observer-independent facts' assumption, which leads to its substitution by LF's AOE. Such an assumption demands for all performed experiments to have observer-independent results which, for the experiment under consideration, implies the existence of a joint distribution, $P(ABCD|XY)$. Moreover, such a joint distribution is demanded to ensure consistency between the outcomes of friends and superobservers when $X,Y =1$.

I maintain, however, that the critique in \cite{LF} seems to depend on a particular, arguably not very natural, reading of an ambiguous element in Brukner's presentation. In particular, they read \cite{Bru1} as demanding all measurements that might be performed to have well-defined values, but it is more natural to read it as demanding all performed experiments to have well-defined values, which seems equivalent to what AOE demands. We also saw that Brukner's derivation of the theorem depends on a hidden assumption, to the effect that the results of Alice and Bob, when they perform a Z measurement, necessarily coincide with the results previously obtained by Charlie and Debbie. It seems, then, that this hidden assumption is identical to the demand of consistency between friends and superobservers imposed by conditions ii and iii in AOE. I conclude that, by adopting the most natural interpretation of Brukner's claims, AOE is exactly equivalent to the conjunction of Brukner's `Observer-independent facts' with his hidden assumption.

All this makes it clear that the AOE assumption contains two logically \emph{independent} parts: 1) the claim that there is a joint  distribution for all results and 2) conditions ii and iii, which have to do with the particular case in which Wigner measures in the same basis as his friend. Moreover, it could be argued that, while \cite{LF} take both parts as demands in order to ensure for observed events to be considered observer-independent or absolute, strictly speaking, it is only the first part that is required to enforce absoluteness of observed events. The issue is that, even when Wigner and his friend measure ``in the same basis'' (e.g., $X,Y=1$), it is never the case that what they are comparing is the result of the exact same measurement  (I use the quotations to emphasize the fact that friends and superobservers perform measurements on different systems so, strictly speaking, the bases cannot be the same). As a result, a demand for absoluteness of observed events does not necessarily imply a demand for these two measurements to yield the exact same result.

The issue is that there are alternative ways to describe a measurement by Wigner, when he measures ``in the same basis'' as his friend. For instance, according to \cite{Bru1}, Alice and Bob might perform a ``Z measurement'' in basis $\{ \ket{Z_+}_i,\ket{Z_-}_i\}$ with $\ket{Z_\pm}_i $ the state of whole lab $i$ after the friend measures the spin along $z$ and finds the result $\pm$. However, according to \cite{LF}, if $X=1$, Alice opens Charlie's vault, asks him what he observed and sets her result equal to that of Charlie. Note also that, when the measurements by Alice and Bob are described in \cite{LF}, they are treated differently, depending on whether they measure ``in the same basis'' than the friend or not. As we saw, when they do, the measurements are described as them simply opening the door. However, when they measure in a different basis, it is argued that the measurement can be thought of as one in which Alice and Bob undo the measurement of the friend, and measure the particle again directly. It is clear that this description is available only by assuming that the measurements are \emph{reversible}, which might not be the case for some models. 

In sum, the AOE assumption contains two independent parts, one demanding for there to be a joint distribution for all results and another demanding consistency between the results of Wigner and his friend when they both measure ``in the same basis''.  And, while it might seem strange to forgo the second, such might be a necessary consequence of certain models, i.e.,  contextual ones (a feature all hidden variable theories must possess). In any case, these two parts are logically independent and it serves us best to make that explicit. Moreover, it might be argued that, while the former does seem to be a necessary demand to make in order to attain objectivity of observed results, the second might be relaxed.\footnote{\cite{Moreno} does exactly that.} 

As I mentioned in section \ref{LF}, an important, far-reaching claim in \cite{LF} is that their theorem ``places strictly stronger constraints on physical reality than Bell's theorem''. To arrive at this claim, first it is argued that, although usually left implicit,  the AOE assumption is necessary for the derivation of Bell's theorem. Then, it is pointed out that, while the derivation of the LF inequalities require AOE, NSD and L, the derivation of the Bell inequalities requires AOE, NSD, L and `Outcome independence'. Since, according to this reasoning, the LF assumptions are a subset of Bell's assumptions, it is concluded that it is logically impossible to construct a model that allows violation of the LF inequality, but does not allow violation of Bell's inequalities.

I contend, though, that, strictly speaking, the AOE assumption is \emph{not} necessary for the derivation of Bell's theorem. Therefore, it is not the case that the LF theorem is strictly stronger than Bell's theorem. As we saw, the AOE assumption involves two independent parts; and, while it seems true that demanding for results to be objective is, in fact, a necessary (implicit) supposition behind Bell's result, a demand of consistency between the results of friends and superobservers, is not required for the derivation of Bell's theorem. In fact, Bell's experimental scenario does not even involve friends and superobservers, so any condition constraining the relation between the results of friends and superobservers, seems fully irrelevant.



The last issue I would like to analyze has to do with the wording of the LF theorem. As we saw in section \ref{LF}, the construction of the theorem has two parts. First, it is shown that models obeying the LF assumptions satisfy the LF inequalities. Second, it is argued for the following conditional: if the quantum operations required by the proposed experiment can, in principle, be performed, then the quantum predictions violate the LF inequalities. Putting everything together, by defining a `physical theory' as ``any theory that correctly predicts the correlations between the outcomes observed by the superobservers Alice and Bob'', the LF theorem is stated as a conditional: if a superobserver can perform arbitrary quantum operations on an observer and its environment, then no physical theory can satisfy the LF assumptions.

I start by inspecting the definition of `physical theory'. As we just saw, in \cite{LF}, that notion is reserved for any theory that ``correctly predicts'' the correlations between Alice and Bob. But what are these ``correct predictions''? Since, as already was established, standard quantum mechanics cannot produce actual predictions for the experiment in question, these ``correct predictions'' cannot be the quantum predictions. Therefore, the ``correct predictions'' must mean those that coincide with the actual, experimentally observed correlations, whatever they turn out to be. Of course, for all we know, these actual correlations can be anything. Therefore, in order to constrain them, \cite{LF} employs a conditional to the effect that, if a superobserver can perform arbitrary quantum operations on an observer and its environment, then the quantum predictions violate the LF inequalities. The idea being that, while, in general, it is true that there are no correct quantum predictions for the proposed experiment, if the antecedent of the conditional is fulfilled, i.e., if a superobserver can, in fact, perform arbitrary quantum operations on an observer and its environment, then unambiguous predictions can be produced. Moreover, the idea is that, in that case, such predictions violate the LF inequalities. The problem is that such a conditional is false.

To begin with, I find the operational tone of the antecedent of the conditional vague and inadequate. The antecedent in question is formulated in \cite{LF} in a couple of different ways, one general and one particular to the EWFS in question. The general form, as we saw, asks for superobservers to be able to perform arbitrary quantum operations on an observer and its environment---that is, for quantum evolution (including quantum measurement) to be controllable on such a scale. The particular formulation demands for the quantum operations required by the proposed experiment to be performable, at least in principle. Either way, what is it exactly that the condition demands? Presumably, the idea is something like this. If, for instance, the measurement of one of the friends objectively breaks unitarity, then the superobserver would not be able to perform a quantum operation that maintains the friend (and her environment) in an entangled superposition corresponding to different observational states. That is, the fact that unitary is broken, is supposed to restrict the operations available to the superobserver.

It seems to me, however, that the fact that a measurement of a friend breaks unitarity is better understood as a feature of the internal dynamics of the model in question, and not really as a constraint on what operations the superobservers can or cannot perform. In other words, even if the measurements of the friends break unitarity, the superobservers are perfectly capable of performing the quantum operations required by the proposed experiment, namely, certain measurements on the corresponding labs---i.e., the breakdown of unitarity does not restrict the measurements available to the superobservers. It seems, then, that what the antecedent of the conditional needs to capture is a constraint on how macroscopic systems evolve during measurements. And, even though \cite{LF} shies away from using these terms, what it seems to be actually demanded is for the evolution during those measurements to be purely unitary.

More importantly, even if the measurements of the friends involve purely unitary evolution, and the superobservers are able to perform quantum operations that maintain the friends in coherent superpositions, that does not mean that the correlations predicted are going to break the LF inequalities. That is, even if the states of the labs, after the friends measure, are given by what quantum mechanics calls a coherent superposition of different observational states, it is not necessarily the case that the LF inequalities would be broken. As we saw already, standard quantum mechanics cannot make predictions for the EWFS in question. Therefore, to make predictions for them, it is necessary to come up with an alternative, non-standard framework, capable of doing so. The point I want to make is that the fact that one of these frameworks stipulates purely unitary evolution during measurements in no way implies that such a model predicts violations of the LF inequalities. As I said, these models must be non-standard alternatives to the standard framework and, as such, they can possess all sorts of non-standard features, including predictions that do not break the LF inequalities, even if superobservers are able to fully control the quantum states of the friends. We must not forget that the assumption that results are objective, together with the assumption of purely unitary evolution, implies the existence of additional variables \citep{Tim}. And these additional variables can play a non-trivial role in the calculation of predictions, allowing for predictions that do not break the LF inequalities.

In sum, \cite{LF} looks for a condition that would guarantee predictions that break the LF inequality and it tries to formulate it in terms of the operations the superobservers are able to perform. The problem is that such a proposal does not work and, more generally, there seems to be no simple condition that would achieve what they are looking for. The truth is that, to make predictions for these scenarios, alternatives to the standard framework are required and different models will make different predictions, without there being a single condition that could determine on which side of the LF inequalities the predictions of a model would land. Of course, what \cite{LF} does show is that models satisfying the LF assumptions do not break the inequalities, but what is lacking is a way to stipulate what models are able to break them.

We can contrast all this with Bell's theorem. In that case, one has 1) that models satisfying Bell's assumptions satisfy Bell's inequality and 2)  that standard quantum mechanics unambiguously predicts violations of the inequality. The theorem can then be paraphrased as ``the predictions of local models are incompatible with those of quantum mechanics''. In the LF case, in contrast, one shows that models satisfying the LF assumptions satisfy the LF inequalities, but one cannot show that quantum mechanics predicts violations of the inequality---and it is not even clear how to characterize models that violate it. One is then only left with the connection between the LF assumptions and the LF inequality. 

Now, it is of course true that, once Bell experiments have been performed, and clear violations of the inequalities have been observed empirically, the fact that standard quantum mechanics violates Bell's inequality becomes almost irrelevant. With experimental violations of the inequality, what we now have is a much stronger claim, namely, that ``the predictions of local models are incompatible with actual experiments''---a claim which is completely independent of quantum mechanics. Couldn't one do the same with the LF inequalities? The issue is that, unlike Bell's experiment, the one required to experimentally probe the LF inequalities, while presumably doable in principle, is unrealizable in practice (and will be so for the foreseeable future). It is true, then, that such experiments would set constraints on empirically viable models, but the fact that we are unable to actually perform them greatly diminishes the force behind the LF result.

Summing up, in \cite{LF} it is shown that models satisfying the LF assumptions satisfy the LF inequalities. However, contrary to what it is argued there, it is not the case that such assumptions are strictly weaker that those of Bell. Moreover, since quantum mechanics is unable to make predictions for the EWFS considered, there is no standard violating the LF inequalities with which to contrast models satisfying the LF assumptions. Finally, the proposed experiment is, at this stage, no more than a gedankenexperiment, so the prospects of experimentally constraining LF models seem, at least for a long time, unattainable.

\section{Conclusions}
\label{Co}

The ``no-go theorem for observer-independent facts'' and the ``Local Friendliness no-go theorem'' are taken by their authors to impose non-trivial constraints on physical reality. However, such theorems suffer from a list of shortcomings that question their validity and limit their strength. In this work, I have shown that the theorem in \cite{Bru1} depends on a non-trivial, implicit assumption, regarding the relation between the results of friends and superobservers, and that it relies on the mistaken notion that standard quantum mechanics is able to make predictions for the EWFS considered.

As for the theorem \cite{LF}, I have shown that, contrary to what is alleged, it does not rest on assumptions that are strictly weaker that those of Bell. Moreover, it relies on the same mistaken application of standard quantum mechanics to the EWFS in question, so there is no standard with which to contrast models satisfying the LF assumptions. Finally, I note that the proposed EWFS is unrealizable in practice, a fact that greatly diminishes the force behind the LF result.

In the end, all these theorems offer is the fact that models satisfying a certain group of assumptions satisfy certain inequalities. However, without theoretical or experimental benchmarks with which to compare those models, the established fact sharply looses impact and relevance. I conclude that the theorems fully fail in imposing significant constraints on the possible nature of physical reality.



\bibliographystyle{apalike}
\bibliography{bibLF.bib}


\end{document}